\documentclass[a4paper,12pt]{article}
\pdfoutput=1
\usepackage{graphicx, rotating,amssymb,enumerate}

\ifx\pdfoutput\undefined
\usepackage[dvips,bookmarks]{hyperref}	
\else
\usepackage{hyperref}	
\fi
\hypersetup{colorlinks,bookmarksopen,bookmarksnumbered,citecolor=verdes,
linkcolor=blus,pdfstartview=FitH,urlcolor=rossos}
\def\myurl#1#2{\href{http://#1}{#2}}
\def\hhref#1{\href{http://arxiv.org/abs/#1}{#1}} 

\usepackage{multicol}
\usepackage{color}
\definecolor{rosso}{cmyk}{0,1,1,0.4}
\definecolor{rossos}{cmyk}{0,1,1,0.55}
\definecolor{rossoc}{cmyk}{0,1,1,0.2}
\definecolor{blu}{cmyk}{1,1,0,0.3}
\definecolor{blus}{cmyk}{1,1,0,0.6}
\definecolor{bluc}{cmyk}{1,1,0,0.1}
\definecolor{verde}{cmyk}{0.92,0,0.59,0.25}
\definecolor{verdec}{cmyk}{0.92,0,0.59,0.15}
\definecolor{verdes}{cmyk}{0.92,0,0.59,0.4}

\font\tenrsfs=rsfs10 at 12pt
\font\sevenrsfs=rsfs7
\font\fiversfs=rsfs5
\newfam\rsfsfam
\textfont\rsfsfam=\tenrsfs
\scriptfont\rsfsfam=\sevenrsfs
\scriptscriptfont\rsfsfam=\fiversfs
\def\mathscr#1{{\fam\rsfsfam\relax#1}}

\newcommand{\GeV}{\,{\rm GeV}}

\def\circa#1{\,\raise.3ex\hbox{$#1$\kern-.75em\lower1ex\hbox{$\sim$}}\,}

\newcommand{\beq}{\begin{equation}}
\newcommand{\eeq}{\end{equation}}

\def\circa#1{\,\raise.3ex\hbox{$#1$\kern-.75em\lower1ex\hbox{$\sim$}}\,}

\newcommand{\antiHe}{\overline{\rm He}}
\newcommand{\He}{\rm He}

\makeatletter

\def\art{\@ifnextchar[{\eart}{\oart}}
\def\eart[#1]#2#3#4#5#6{{\rm #2}, {#3 #4} {\rm (#6) #5} [{\hhref{#1}}]}
\def\hepart[#1]#2{{\rm #2, \hhref{#1}}}
\newcommand{\oart}[5]{{\rm #1}, {#2 #3} {\rm (#5) #4}}

\newcounter{alphaequation}[equation]
\def\thealphaequation{\theequation\hbox to
0.6em{\hfil\alph{alphaequation}\hfil}}
\def\eqnsystem#1{
\def\@eqnnum{{\rm (\thealphaequation)}}
\def\@@eqncr{\let\@tempa\relax \ifcase\@eqcnt \def\@tempa{& & &} \or
  \def\@tempa{& &}\or \def\@tempa{&}\fi\@tempa
  \if@eqnsw\@eqnnum\refstepcounter{alphaequation}\fi
\global\@eqnswtrue\global\@eqcnt=0\cr}
\refstepcounter{equation} \let\@currentlabel\theequation \def\@tempb{#1}
\ifx\@tempb\empty\else\label{#1}\fi
\refstepcounter{alphaequation}
\let\@currentlabel\thealphaequation
\global\@eqnswtrue\global\@eqcnt=0 \tabskip\@centering\let\\=\@eqncr
$$\halign to \displaywidth\bgroup \@eqnsel\hskip\@centering
$\displaystyle\tabskip\z@{##}$&\global\@eqcnt\@ne
\hskip2\arraycolsep\hfil${##}$\hfil& \global\@eqcnt\tw@\hskip2\arraycolsep
$\displaystyle\tabskip\z@{##}$\hfil
\tabskip\@centering&\llap{##}\tabskip\z@\cr}
\def\endeqnsystem{\@@eqncr\egroup$$\global\@ignoretrue} \makeatother

\def\AMS{{\sc Ams-02}}

\begin{document}
\begin{flushright}
\footnotesize
SACLAY--T14/003
\end{flushright}
\color{black}

\begin{center}


{\LARGE\bf Anti-helium from Dark Matter annihilations}

\medskip
\bigskip\color{black}\vspace{0.5cm}

{
{\large\bf Marco Cirelli}\ $^a$,
{\large\bf Nicolao Fornengo}\ $^{b,c}$,\\[3mm]
{\large\bf Marco Taoso}\ $^{a}$,
{\large\bf Andrea Vittino}\ $^{a, b,c}$
}
\\[7mm]
{\it $^a$ \href{http://ipht.cea.fr/en/index.php}{Institut de Physique Th\'eorique}, CNRS, URA 2306 \& CEA/Saclay,\\ 
	F-91191 Gif-sur-Yvette, France}\\[3mm]
{\it $^b$ \href{http://www.unito.it/unitoWAR/appmanager/dipartimenti8/D076?_nfpb=true}{Department of Physics}, University of Torino,\\ via P. Giuria 1, I-10125 Torino, Italy}\\[3mm]
{\it $^c$ \href{http://www.to.infn.it}{INFN} - Istituto Nazionale di Fisica Nucleare, Sezione di Torino,\\ via P. Giuria 1, I-10125 Torino, Italy}
\end{center}

\bigskip

\centerline{\large\bf Abstract}
\begin{quote}
\color{black}
Galactic Dark Matter (DM) annihilations can produce cosmic-ray anti-nuclei via the nuclear coalescence of the anti-protons and anti-neutrons originated directly from the annihilation
process. Since anti-deuterons have been shown to offer a distinctive DM signal, with potentially
good prospects for detection in large portions of the DM-particle parameter space, we explore
here the production of heavier anti-nuclei, specifically anti-helium. Even more than for anti-deuterons,
the DM-produced anti-He flux can be mostly prominent over the astrophysical anti-He background at low kinetic energies, typically below 3-5 GeV/n. However, the larger number of anti-nucleons involved in the formation process makes the anti-He flux extremely small.
We therefore explore, for a few DM benchmark cases, whether the yield is sufficient to allow for anti-He detection in current-generation experiments, such as \AMS. 
We account for the uncertainties due to the propagation in the Galaxy and to the uncertain details of the coalescence process, and we consider the constraints already imposed by anti-proton searches. We find that only for very optimistic configurations might it be possible to achieve detection with current generation detectors.
We estimate that, in more realistic configurations, an increase in experimental sensitivity at low kinetic energies of about a factor of 500-1000 would allow to start probing DM through the rare cosmic anti-He production.
\end{quote}


\section{Introduction}
\label{sec:introduction}

The particle Dark Matter (DM) which is believed to constitute the halo of our Galaxy (as well as shaping the large scale structures of the Universe) is proving to be more elusive than ever to direct searches. The search for possible `smoking guns' is therefore as important as it has ever been. One of such smoking gun could be the observation of exotic light anti-nuclei in the cosmic radiation, possibly produced by DM via the rapid coalescence of the anti-nucleons ($\bar p$ and $\bar n$) emerging among the final products of the annihilation processes in the galactic halo. Indeed, the astrophysical background for these species is predicted to be extremely reduced and to be peaking in a range of energies typically different from the one of the DM-originated one (this is due essentially to the different kinematics in the production processes, as we will recall below).
Hence, the claim is often made that the detection of even a single anti-nucleus in the energy range predicted for DM could constitute a very compelling hint in favor of DM, or at least of a very exotic process different from spallations of cosmic rays.
A well explored scenario is the one of anti-deuterons, which have been proposed more than a decade ago~\cite{Dbarhistoric} and continue to be of interest~\cite{Dbarrecent,Kadastik:2009ts,ourantiD}. 
In this paper, instead, we ask whether anti-helium (hereafter anti-He) nuclei could be produced in DM annihilations with a sizable yield to which current experiments such as \AMS\ could be sensitive. Indeed, the \AMS\ experiment lists among its physics goals the search for anti-He and it foresees to push the sensitivity down by several orders of magnitude with respect to the bounds imposed by other recent experiments~\cite{AMSantiHe}. It is therefore timely to investigate whether a signal from DM could emerge in such an exotic channel. 

\medskip

The rest of this short paper is organized as follows. In Sec.~\ref{sec:coalescence} we review the production mechanism of anti-He, while in Sec.~\ref{sec:propagation} we review its propagation in the Galaxy. In Sec.~\ref{sec:results} we present the predicted fluxes for a few relevant benchmark DM cases and in Sec.~\ref{sec:conclusions} we put forward our conclusions.


\section{Production by coalescence}
\label{sec:coalescence}

The production of anti-nuclei in a given reaction is usually described within the framework of the so-called coalescence model~\cite{coalescence,Duperray}. The idea behind this approach is very simple: some of the anti-nucleons produced in the reaction under scrutiny can merge to form an anti-nucleus if their relative momenta is less than an effective parameter, the coalescence momentum $p_{\rm coal}$, which is usually determined from comparison with experimental data (when available).  By following the coalescence approach, the spectrum of an anti-nucleus $\bar{A}$ with mass number $A$ can be written as:   
\begin{equation}
\gamma_{\bar{A}}\frac{d^3N_{\bar{A}}}{d^3k_{\bar{A}}} = \left(\frac{4}{3}p_{\rm coal}^3\right)^{A-1}\left(\left.\gamma_{\bar{p}}\frac{d^3N_{\bar{p}}}{d^3k_{\bar{p}}}\right|_{k_{\bar{p}}=k_{\bar{A}}/A}\right)^A
\label{eq:coalescence}
\end{equation} 
As one should expect, the coalescence mechanism predicts that the increase of the mass number $A$ comes with a rapid growth in the suppression factor for the yield of the anti-nucleus $\bar{A}$. As a rule of thumb (based on the results of \cite{Duperray,Salati} for the background component and of~\cite{PPPC4DMID,ourantiD} for the $\bar p$ and $\bar d$ fluxes from DM), we estimate that for each additional anti-nucleon involved in the merging process one would have a decrease of the yield by a factor ${\cal O}$($10^{-4}$). This can also be seen from the coalescence formula above, plugging in typical numbers for the parameters ($p_{\rm coal} \sim \mathcal{O}(0.1)$ in GeV and knowing that the $\bar p$ spectrum, normalized in terms of particles per annihilation event, peaks at $\mathcal{O}(10^{-1})$ for the kinetic energies of few GeV at most in which we will be interested).
Thus, we decide to focus only on the anti-$^3$He 
and to disregard completely the contribution from anti-$^4$He.

\medskip
 
As shown in~\cite{Kadastik:2009ts} for the case of the anti-deuteron production, in order to have a correct computation of the anti-nuclei yields, the details of the angular distribution of the anti-nucleons in the final state, together with possible (anti-)correlations between them, must be taken into account. This can be done by using a MonteCarlo (MC) coalescence model which basically consists in checking on an {\it event-by-event} basis if the anti-nucleons that are produced in a DM annihilation event (which is simulated by using a MC event generator) are sufficiently close in momentum space for the coalescence to occur. In this paper we adopt this approach: for our MC coalescence model we use the MC event generator {\sc Pythia} 6.4.26~\cite{Pythia6.4} and, for each DM candidate that we consider, 
we simulate ${\cal O}$($10^{10}$) annihilation events (the exact number depending on
the condition of reaching a sufficient anti-He statistics).  We assume that three anti-nucleons ${N_1,N_2,N_3}$ merge in a single bound state if all their relative momenta 
($\| \vec{k}_{N_1}-\vec{k_{N_2}}\|,\|\vec{k}_{N_2}-\vec{k}_{N_3}\|,\|\vec{k}_{N_1}-\vec{k}_{N_3}\|$) 
are smaller than $p_{\rm coal}$, being $\vec{k}_{N_i}$ the anti-nucleon momenta in the center of mass frame of the ${N_1,N_2,N_3}$ system (which corresponds to the rest frame of the bound state).
Other prescriptions would be possible, as long as a `first-principles' description of coalescence is not available. For instance, one could impose that the three momenta lie inside a minimum bounding sphere in momentum space with a diameter determined by $p_{\rm coal}$, as done in~\cite{Carlson:2014ssa}. However, we have checked explicitly~\footnote{We acknowledge private communications with Eric Carlson on this point.} that the two methods differ in the determination of the required momentum at most by 15\%, which, in light of the larger uncertainties that we will discuss below, can be neglected.\\
In addition, one can easily understand that the anti-He spectrum can be overestimated if the information about the anti-nucleons positions in the physical space is completely disregarded: in fact, it is highly unlikely to have a coalescence if the three anti-nucleons are formed far from each other (as it is if, for example, one of them comes from the decay of a relatively long-lived particle) \cite{ourantiD}. This condition is taken into account by switching off (to the maximal extent allowed by the default setup of the event generator) the decay of all the long-lived particles ({\it i.e.} those with a lifetime $\tau > 10^{-15}$ sec) in our MC event generator. Imposing a more stringent restriction on the position (e.g. based on the size of a helium nucleus, some fm) would be extremely time consuming from the point of view of the numerical running and would not actually have an important impact on the results. Indeed, it has been shown in detail in~\cite{ourantiD} that the reduction of the flux caused by the size constraint amounts at most to 30\% (with the exception of heavy quark channels from light DM, which, for this reason, we will not choose as a benchmark case in the following).

\medskip

  The anti-nucleons that can take part in the formation process of an anti-He nucleus can be either two $\bar{p}$ and one $\bar{n}$ (and in this case the anti-He is formed directly) or two $\bar{n}$ and one $\bar{p}$ (i.e. in this case the anti-He is the result of the formation of an anti-tritium that subsequently decays into an anti-He in a process that, given the typical propagation scales with which we are dealing, can be considered as occurring instantaneously). However, as stated in \cite{Salati}, we expect the direct formation of the anti-He in the $\bar{p} \bar{p} \bar{n}$  channel to be suppressed by Coulombian repulsion between the two anti-protons. Such repulsion could also induce spectral distortions. 
Thus, in the following, in order to be as conservative as possible, we will only show the anti-He yields that are produced by the coalescence in the $\bar{p} \bar{n} \bar{n}$ channel. However, we checked that, if the same coalescence momentum is used for the two cases, these two contributions are practically equal for all the benchmark cases that we consider (see Section \ref{sec:results}), this being an expected consequence of the fact that the $\bar{p}$ and the $\bar{n}$ production cross sections, in a DM annihilation event, are almost equal. Thus, if one wants to add also the contribution from the coalescence in the $\bar{p} \bar{p} \bar{n}$ channel to the anti-He yields that we show in Section \ref{sec:results}, it is sufficient to multiply the fluxes by a factor 2 (if the Coulombian repulsion is completely neglected) or smaller (if the Coulombian repulsion is taken into account). 
  
\medskip
     
Experimental data on the anti-He (or anti-tritium) production are extremely scarce in the literature and they refer uniquely to proton-nucleus \cite{antiHedatapA} or heavy-ions collisions \cite{antiHedataAA} whose dynamics is clearly very different from the one of a DM pair annihilation reaction. Thus, we decide to use as a reference value for the coalescence momentum  $p_{\rm coal}$ the one that was found in \cite{ourantiD} to reproduce, within the same MC coalescence algorithm, the {\it anti-deuteron} production rate in $e^+e^-$ collisions at the $Z$ resonance measured by the {\sc Aleph} collaboration at the LEP collider~\cite{Aleph}, {\it i.e.} $p_{\rm coal} = 195$ MeV. This is of course a somewhat arbitrary choice, but it is supported by the findings of ref.~\cite{Duperray}, which shows (for the case of astrophysical spallations) that adopting the same coalescence momentum for anti-D and anti-He leads to consistent results. However, we must warn the reader that the value of the $p_{\rm coal}$ parameter largely affects our results, as its clear also from Eq. (\ref{eq:coalescence}) in which for $A=3$, the dependence from $p_{\rm coal}$ is in the form $p_{\rm coal}^6$. To give an idea of the role played by $p_{\rm coal}$ within our MC coalescence mechanism, in Section \ref{sec:results} we will show how the anti-He flux varies if values of the $p_{\rm coal}$ parameter greater than our reference value are chosen.


\section{Propagation in the Galaxy}
\label{sec:propagation}

Once anti-He nuclei are created at any given point in the galactic halo, they have to propagate through the Galaxy all the way to the collection point (the Earth). 
The suitable formalism to follow this process resembles closely the one adopted for anti-protons or anti-deuterons, reviewed e.g. in~\cite{PPPC4DMID}, to which we refer for further details and references. We here only summarize the main points. 

The propagation of charged nuclei is described by a differential equation incorporating the different processes that they undergo: 
\beq 
\label{eq:diffeq}
\frac{\partial f}{\partial t} - \mathcal{K}(T)\cdot \nabla^2f + \frac{\partial}{\partial z}\left( {\rm sign}(z)\, f\, V_{\rm conv} \right) = Q-2h\, \delta(z)\, \Gamma f .
\eeq
Here $f(t,\vec x,T) = dN_{\mbox{\tiny He}}/dT$ is the number density of anti-He nuclei per unit kinetic energy $T$, in a given location $\vec x$ and at a given time $t$. $\mathcal{K}(T) = \mathcal{K}_0 \beta \, (p/\GeV)^\delta$ is the coefficient of the process of diffusion of the anti-nuclei on the magnetic field inhomogeneities (with $p = (T^2 +2 m_{_{\antiHe}} \,T)^{1/2}$ and $\beta = v_{_{\antiHe}}/c = \left(1-m_{_{\antiHe}}^2/(T+m_{_{\antiHe}})^2\right)^{1/2}$ the anti-nucleus momentum and velocity).
$V_{\rm conv}$ is the velocity of the galactic convective wind. The quantity:
\begin{equation}
Q = \frac{1}{2} \left(\frac{\displaystyle \rho}{m_{\rm DM}}\right)^2 \sum_{\alpha} \langle \sigma v\rangle_\alpha \ \frac{dN^{^\alpha}_{_{\antiHe}}}{dT} 
\end{equation}
represents the source term due to DM annihilations (with thermally averaged cross section $\langle \sigma v\rangle$), summed over the different channels $\alpha$. Several different profiles can be considered for the DM density $\rho$: Navarro-Frenk-White (denoted `NFW'), Moore (`Moo'), Isothermal (`Iso'), Einasto (`Ein'), Burkert (`Bur') and contracted Einasto (`EiB'). We refer to~\cite{PPPC4DMID} for their precise definitions in terms of functional forms and parameters. 
The last term describes the interactions of anti-He on the interstellar gas in the galactic plane
(with a thickness of $h=0.1\,{\rm kpc}$) with rate
$\Gamma = (n_{\rm H} + 4^{2/3} \, n_{\rm He}) \ \sigma_{p-\antiHe} \ v_{_{\antiHe}}$,
where $n_{\rm H}\approx 1/{\rm cm}^3$ is the disk hydrogen density and $n_{\rm He}\approx 0.07\, n_{\rm H}$ is the disk helium density (the factor $4^{2/3}$ accounting for the different geometrical cross section in an effective way). For the nuclear cross sections we use the parametrizations in Table 4.5 of~\cite{DuperrayThese}. 

\begin{table} 
\begin{center}{\small
\begin{tabular}{c|cccc}
 &  \multicolumn{4}{c}{Galactic charged CR propagation parameters}  \\
Model  & $\delta$ & $\mathcal{K}_0$ [kpc$^2$/Myr] & $V_{\rm conv}$ [km/s] & $L$ [kpc]  \\
\hline 
{\sc Min}  &  0.85 &  0.0016 & 13.5 & 1 \\
{\sc Med} &  0.70 &  0.0112 & 12 & 4  \\
{\sc Max}  &  0.46 &  0.0765 & 5 & 15 
\end{tabular}}
\caption{\em \small {\bfseries Propagation parameters} in the galactic halo (from~\cite{DonatoPRD69}). 
\label{tab:proparam}}
\end{center}
\end{table}

\begin{figure}[!t]
\begin{minipage}{0.6\textwidth}
\tiny{
\begin{tabular}{c|c|rrrrrr}
\multicolumn{8}{l}{\footnotesize DM annihilation} \\
\hline
\rm{halo} & \rm{prop} & $a_0$ & $a_1$ & $a_2$ & $a_3$ & $a_4$ & $a_5$ \\[0.3mm]
\hline
  & {\sc Min} & 0.5019 & 0.5278 & -0.2395 & -0.0493 & 0.0197 & -0.0017 \\
 \rm{NFW} & {\sc Med} & 1.3061 & 0.3998 & -0.1698 & -0.0158 & 0.0047 & -0.0001 \\
  & {\sc Max} & 2.0432 & 0.0012 & -0.0477 & 0.0046 & -0.0052 & 0.0008 \\
  \hline
  & {\sc Min} & 0.5019 & 0.5278 & -0.2395 & -0.0493 & 0.0197 & -0.0017 \\
 \rm{Moo} & {\sc Med} & 1.3226 & 0.4248 & -0.1523 & -0.0233 & 0.0054 & -0.0001 \\
  & {\sc Max} & 2.1162 & 0.0251 & -0.0490 & 0.0032 & -0.0049 & 0.0008 \\
  \hline
  &  {\sc Min} & 0.5019 & 0.5278 & -0.2395 & -0.0493 & 0.0197 & -0.0017 \\
 \rm{Iso} & {\sc Med} & 1.2857 & 0.3626 & -0.1923 & -0.0115 & 0.0060 & -0.0004 \\
  & {\sc Max} & 1.9253 & -0.0381 & -0.0469 & 0.0064 & -0.0053 & 0.0008 \\
  \hline
  &  {\sc Min} & 0.5019 & 0.5278 & -0.2395 & -0.0493 & 0.0197 & -0.0017 \\
 \rm{Ein} & {\sc Med} & 1.3388 & 0.3704 & -0.1485 & 0.0022 & -0.0053 & 0.0011 \\
  & {\sc Max} & 2.1354 & 0.0054 & -0.0375 & -0.0044 & -0.0026 & 0.0006 \\
  \hline
  &  {\sc Min} & 0.5019 & 0.5278 & -0.2395 & -0.0493 & 0.0197 & -0.0017 \\
 \rm{EiB} & {\sc Med} & 1.3941 & 0.4175 & -0.1086 & -0.0279 & 0.0023 & 0.0004 \\
  & {\sc Max} & 2.3274 & 0.0006 & -0.0082 & -0.0246 & 0.0028 & 0.0001 \\
  \hline
  &  {\sc Min} & 0.5019 & 0.5278 & -0.2395 & -0.0493 & 0.0197 & -0.0017 \\
 \rm{Bur} & {\sc Med} & 1.2465 & 0.3200 & -0.1975 & -0.0080 & 0.0058 & -0.0004 \\
  & {\sc Max} & 1.8276 & -0.0558 & -0.0472 & 0.0069 & -0.0051 & 0.0008 \\
 \end{tabular}}
\caption{\em \small {\bfseries Propagation function for anti-$^3$He nuclei from annihilating DM}, for the different halo profiles and sets of propagation parameters, and the corresponding fit parameters to be used in eq.~$(\ref{eq:fitantihelium3})$.}
\label{fig:RfunctionsHe3}
\end{minipage}
\quad
\begin{minipage}{0.37\textwidth}
\vspace{-0.9cm}
\centering
\includegraphics[width=\textwidth]{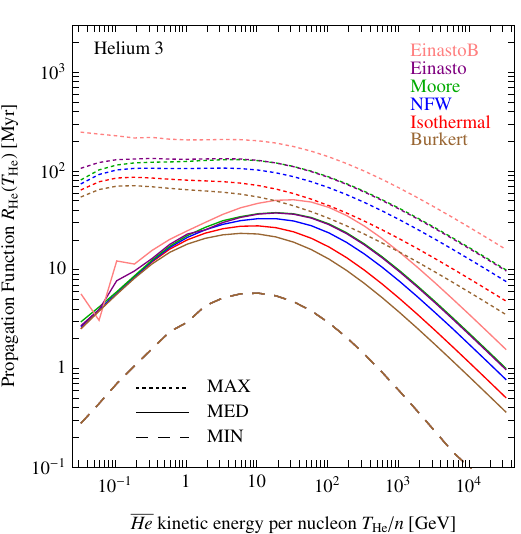}
\end{minipage}
\end{figure}

Since one assumes steady state conditions, the equation is solved as $\partial f/\partial t = 0$. It is solved inside a cylindrical volume with borders $z=\pm L$ and $r=R_{\rm gal} = 20$ kpc (the radius of the Galaxy), on which the particle number density is taken to be vanishing. The propagation parameters entering the formalism are therefore: the normalization of the diffusion coefficient $\mathcal{K}_0$, its power index $\delta$, the velocity $V_{\rm conv}$ and the thickness of the diffusive region $L$. As customary, we consider the sets `{\sc Min}, {\sc Med}, {\sc Max}' as listed in Table~\ref{tab:proparam}. 
The solution for the anti-He differential flux at the position of the Earth $ d\Phi_{_{\antiHe}}/dT\,(T,\vec r_\odot) = v_{_{\antiHe}}/(4\pi) f $ can be cast in a simple factorized form:
\beq
\label{eq:solution}
\frac{d\Phi_{_{\antiHe}}}{dT}(T,\vec r_\odot) = \frac{v_{_{\antiHe}}}{4\pi}  \left(\frac{\rho_\odot}{m_{\rm DM}}\right)^2 R(T)   \sum_\alpha \frac{1}{2} \langle \sigma v\rangle_\alpha \frac{dN^{^\alpha}_{_{\antiHe}}}{dT} 
\eeq
The function $R(T)$ encodes all the astrophysics of production and propagation. There is such a `propagation function' for any choice of DM galactic profile and for any choice of set of propagation parameters among those in Table~\ref{tab:proparam}. 
We explicitely provide $R(T)$ for all these cases in terms of an interpolating function:
\beq
{\rm log}_{10}\left[R(T)/{\rm Myr}\right] = a_0 + a_1\, \kappa + a_2\,  \kappa^2 + a_3\,  \kappa^3 + a_4\,  \kappa^4 + a_5\,  \kappa^5,
\label{eq:fitantihelium3}
\eeq
with $\kappa = \log_{10}T/\GeV$ and the coefficients reported in the table in Fig.~\ref{fig:RfunctionsHe3}. As could be expected, the functions are very similar in shape to the ones relevant for anti-protons and anti-deuterons, presented e.g. in ~\cite{PPPC4DMID}. 

\medskip

The final step consists in applying the effects of the transport of the charged nuclei inside the heliosphere (solar modulation). The details of this process depend on the properties of solar activity (intensity and orientation of the solar magnetic field) at the time of observations, which is of course unknown. We follow the standard formalism (see e.g.~\cite{PPPC4DMID}) and adopt a Fisk potential of 500 MV, which corresponds to a minimum of the solar activity and therefore minimizes the impact of solar modulation on the predictions.


\section{Results and Discussion}
\label{sec:results}

\begin{figure}[p]
\begin{center}
\vspace{-1.5cm}
\includegraphics[width=0.47\textwidth]{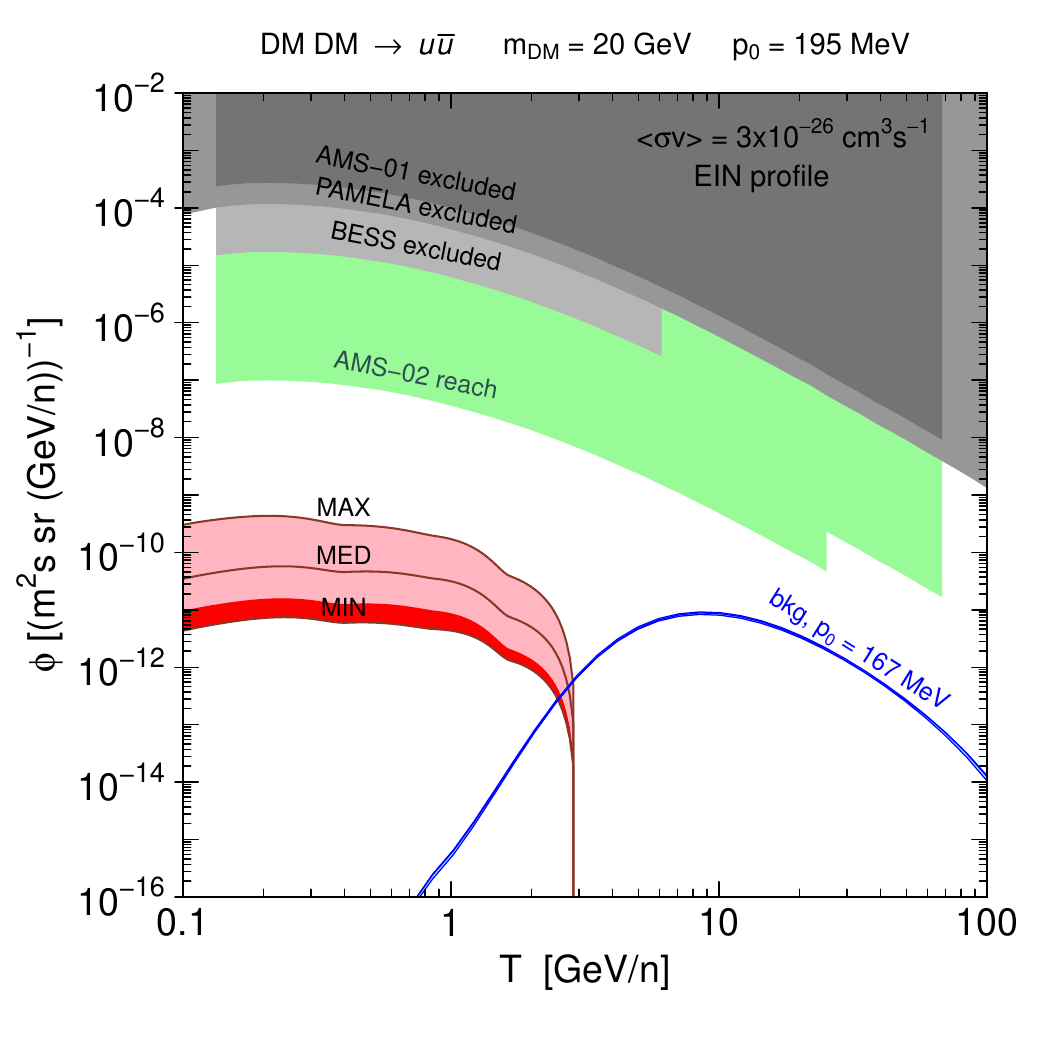}
\includegraphics[width=0.47\textwidth]{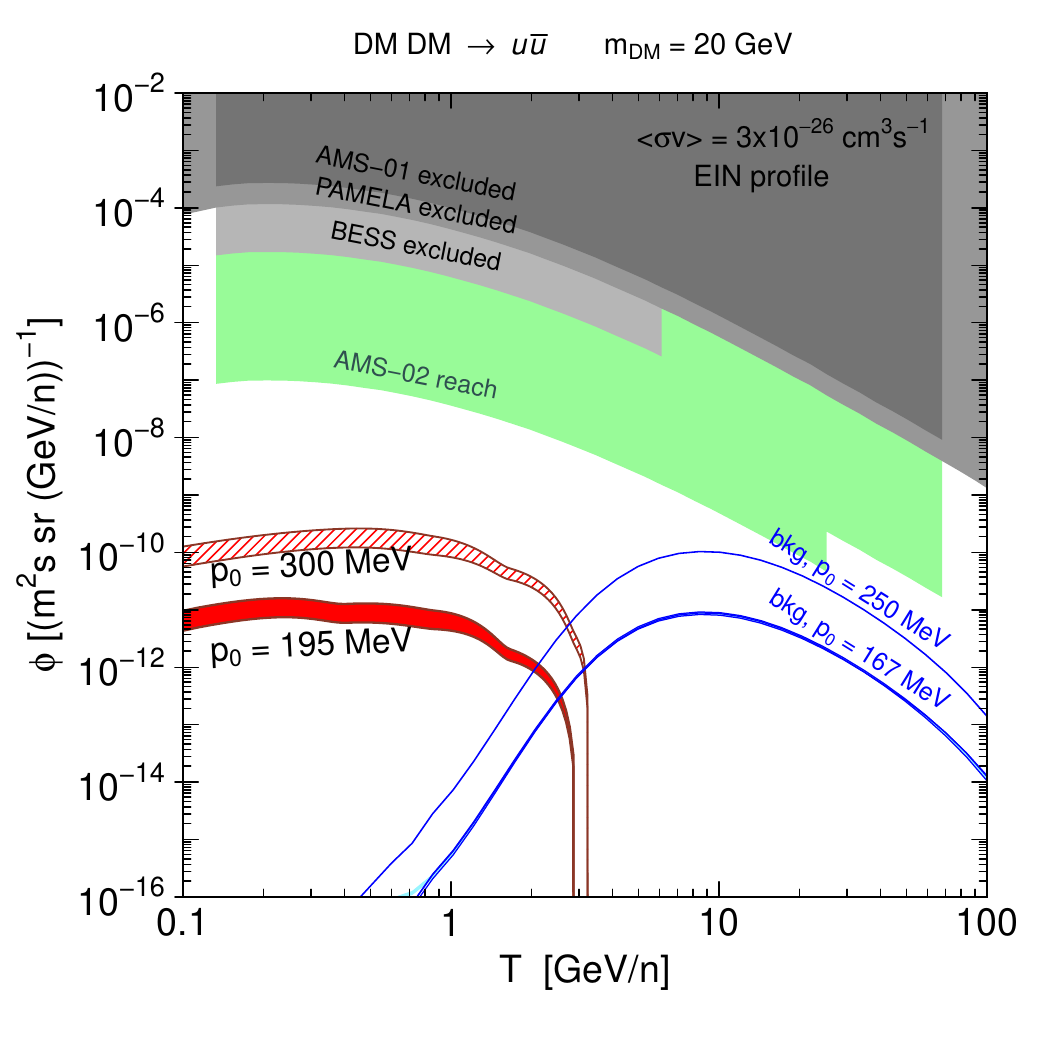} \\
\includegraphics[width=0.47\textwidth]{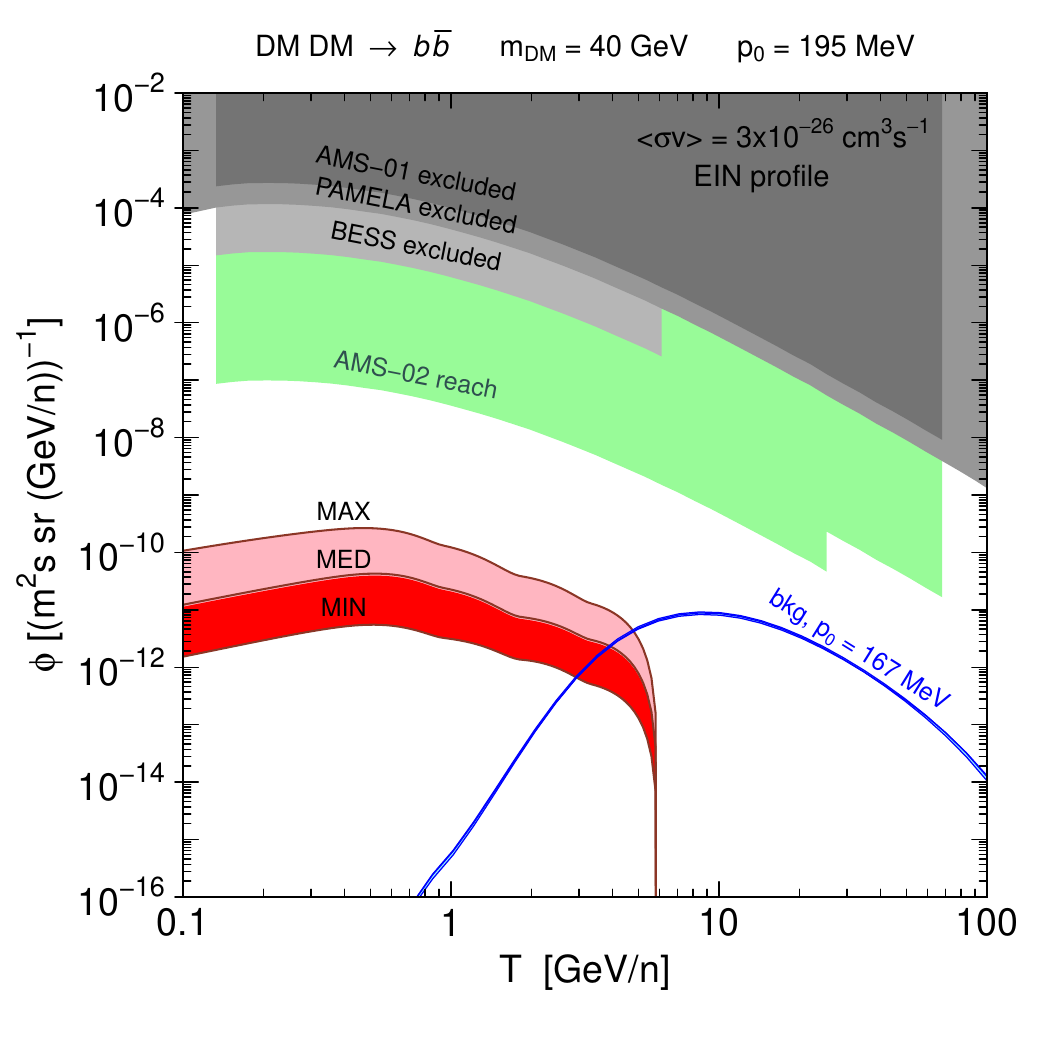}
\includegraphics[width=0.47\textwidth]{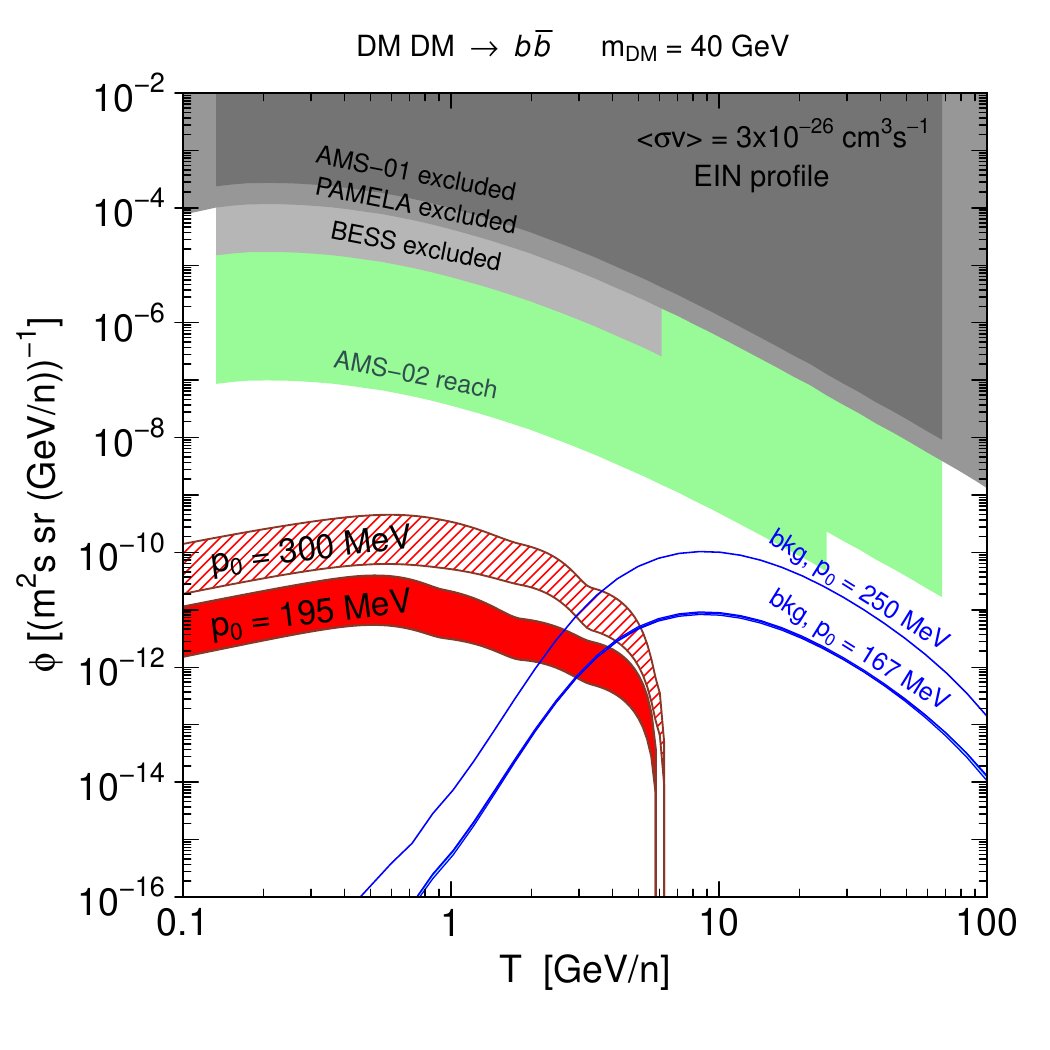} \\
\includegraphics[width=0.47\textwidth]{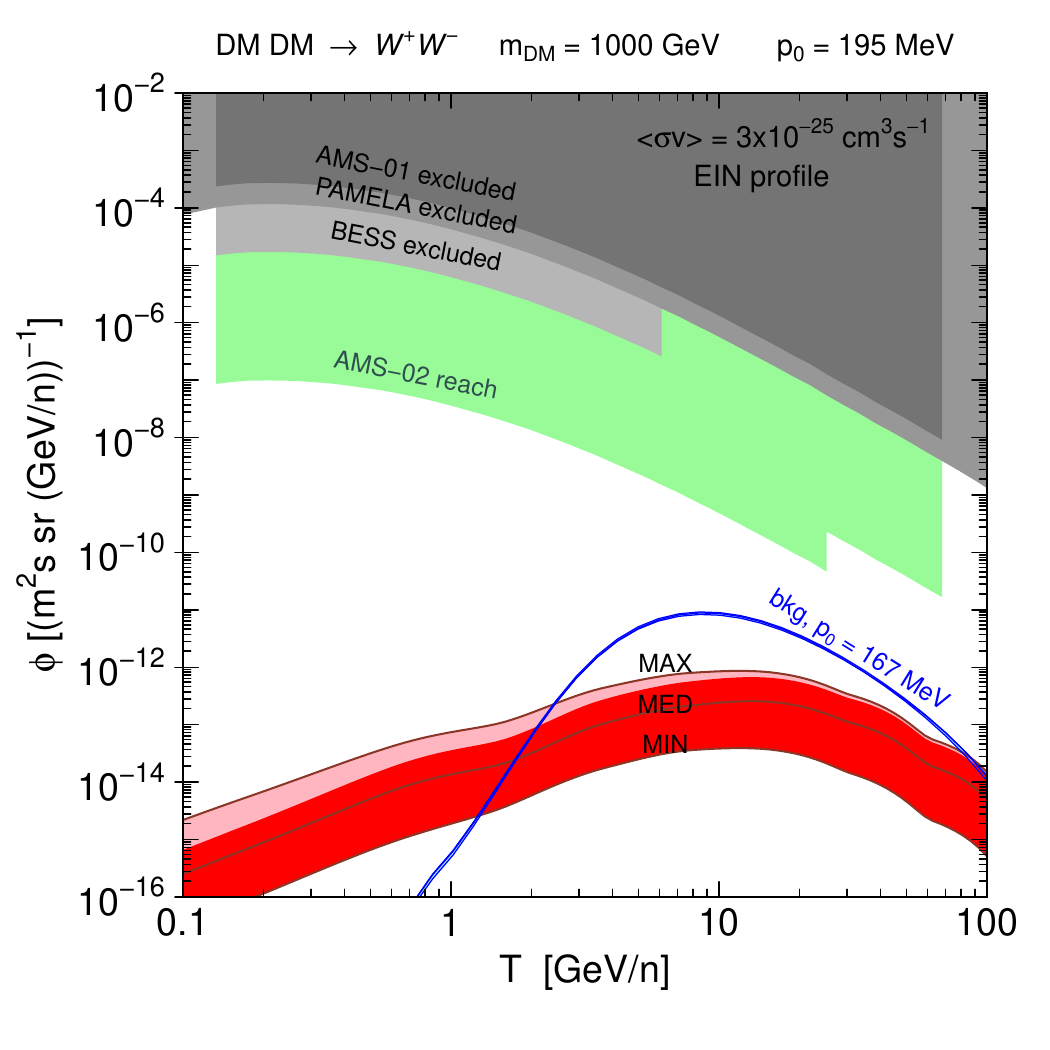}
\includegraphics[width=0.47\textwidth]{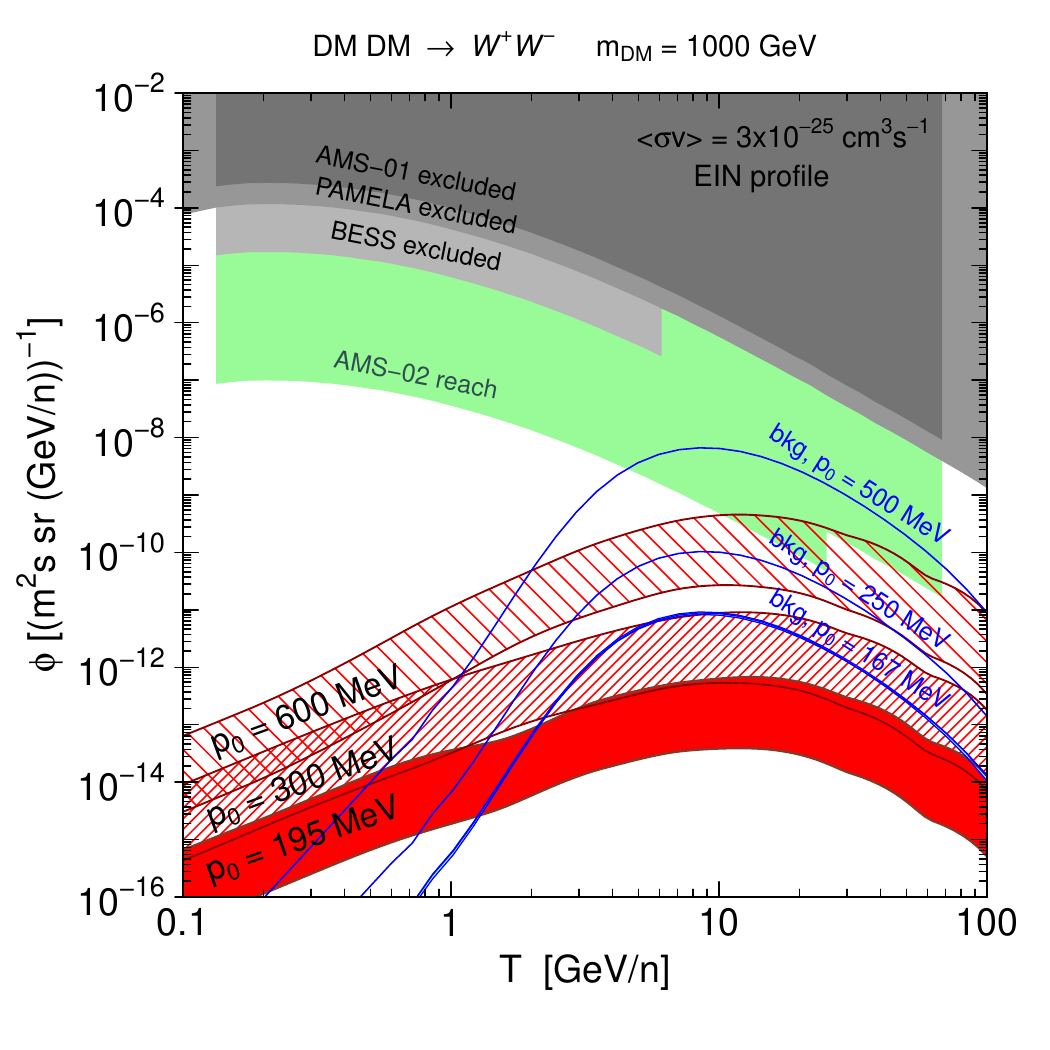}
\caption{\small\em\label{fig:results} {\bfseries Predicted fluxes of anti-He} from the annihilation of a 20 GeV DM particle into light quarks (top row), 40 GeV into $b \bar b$ (middle row) and 1 TeV into $W^+W^-$ (bottom row), compared to the predicted astrophysical background, to the current bounds and to the expected sensitivity of {\sc Ams-02}. Left column: the three lines from bottom to top correspond to {\sc Min}, {\sc Med}, {\sc Max}. Lighter shades individuate fluxes disfavored by $\bar p$ constraints. Right column: varying the coalescence momentum. For the background, the three lines for $p_0 = 167$ MeV (barely distinguishable) also correspond to {\sc Min}, {\sc Med}, {\sc Max}.}
\end{center}
\end{figure}

By folding the production fluxes presented in Sec.~\ref{sec:coalescence} with the propagation functions of Sec.~\ref{sec:propagation} as described in Eq.~(\ref{eq:solution}) we obtain the predicted anti-He spectra from DM annihilation.
We illustrate the results focussing on three benchmark cases: 
\begin{enumerate}[i)]
\item annihilation into light quarks ($u \bar u$ for definiteness) of a 20 GeV DM particle, with thermal annihilation cross section  $\langle \sigma v\rangle = 3 \times 10^{-26}\ {\rm cm}^3/{\rm s}$.
\item annihilation into $b \bar b$ of a 40 GeV DM particle, with thermal annihilation cross section  $\langle \sigma v\rangle = 3 \times 10^{-26}\ {\rm cm}^3/{\rm s}$.
\item annihilation into $W^+W^-$ of a 1 TeV DM particle, with a larger annihilation cross section  $\langle \sigma v\rangle = 3 \times 10^{-25}\ {\rm cm}^3/{\rm s}$.
\end{enumerate}
These span a variety of relevant possibilities: they range on two decades in masses (covering the typical WIMP scale) and they exemplify the channels in which hadronic production is present (light quarks, heavy quarks and gauge bosons). Other channels yield fluxes that are very similar to one of these ($ZZ$ to $W^+W^-$, other heavy quarks to $b\bar b$ and other light quarks to $u \bar u$).
In Fig.~\ref{fig:results} (left) we show the fluxes (dark red lines) for {\sc Min}, {\sc Med} and {\sc Max} propagation parameters.
For what concerns the choice of the DM galactic profile, we choose Einasto for definiteness. Making another choice would have an impact on the prediction that can be easily judged from the span in the propagation functions shown in Fig. \ref{fig:RfunctionsHe3}. Namely, essentially no impact for {\sc Min}, a factor of $\sim$5 for {\sc Max}. 

\medskip

An important point to consider is that the same annihilation process that produces anti-He of course also produces anti-protons, which are tightly constrained~\cite{pbarconstraints} by the {\sc Pamela} measurement~\cite{PAMELApbar} of a spectrum very well consistent with the predicted astrophysical background. We take these constraints into account by disfavoring the portion of the predicted region that is excluded by anti-protons (shaded in lighter color in the left panels of Fig.~\ref{fig:results}). For a concrete example: a model predicting annihilations of a 20 GeV DM particle into light quarks with thermal cross section (the case of the top left panel of Fig.~\ref{fig:results}) is allowed by anti-proton constraints only if the propagation parameters yield a flux somewhere in between {\sc Min} and {\sc Med}~\cite{pbarconstraints}; we therefore shade away the upper portion of the area spanned by the spectra. In practice, we determine the maximal annihilation cross section allowed for {\sc Med} by anti-proton constraints and we then rescale the {\sc Med} anti-He spectrum by the ratio of such cross section and the thermal one. The rescaled spectrum delimitates from above the allowed region (darker red in Fig.~\ref{fig:results}).

In the figures we also show the estimate of the astrophysical background (blue lines), that we discuss in the Appendix. As anticipated, and in analogy with the case of anti-deuterons, the astrophysical spectrum peaks in an energy range that is higher than the one of the DM fluxes (except for the case of a large DM mass). This is essentially due to the different kinematics with which an anti-He nucleus is produced in the astrophysical environment (spallation of high energy cosmic rays on interstellar gas at rest) with respect to the case of DM (annihilation at rest of two heavy particles), as we now proceed to explain (following the discussion in~\cite{Gaisser} of the completely analogous case of anti-protons).
 The astrophysical background is produced by spallations of cosmic-rays protons (and, much less importantly, other nucleons) with the interstellar gas. The minimum incident proton energy that allows an anti-He to be produced is equal to 31 $m_p$: this is implied by conservation of energy and momentum, taking into account the fact that at least five nucleons have to be produced in addition to the anti-He nucleus, in order to conserve the electric charge and the baryon number.
At this threshold, in the center of mass frame, the particles produced in the reaction are at rest and thus the corresponding energy of the produced anti-He nucleus in the laboratory frame is 4 GeV per nucleon. Anti-He nuclei with energies smaller than this can only be produced in configurations where, in the center of mass frame, the anti-He has a momentum with a component aligned in the opposite direction with respect to the momentum of the incident proton. This requires larger energies of the initial proton. Since the cosmic-rays proton flux is a steeply falling function of the energy, this implies, somewhat counterintuitively, a suppression of the anti-He spectrum at energies below the critical one of 4 GeV per nucleon.
For the case of DM annihilations, this kinematical suppression described above is not present. The peak of the anti-He energy spectrum is instead related to the DM mass. In particular the flux of anti-nuclei produced by annihilations of light DM particles, is maximized at energies below few GeV per nucleon, therefore in an energy range where the astrophysical background is suppressed.


\medskip

We show in grey the areas currently excluded by the experiments which have looked for a flux of anti-He in cosmic rays: {\sc Ams-01}~\cite{AMS01bound}, {\sc Bess}~\cite{BESS} and {\sc Pamela}~\cite{PAMELAbound}. 
Since all these experimental results are given in terms of $\antiHe/\He$ ratios, we convert them into anti-He fluxes using the $\He$ flux measured by {\sc Pamela}~\cite{PAMELAHe}\footnote{We correct however the {\sc Pamela} spectrum in order to account for the different value of the solar modulation parameter that we are using here.} ({\sc Ams-02} has also released preliminary data~\cite{AMSHeICRC2013}, that we do not use).
Finally, we show in green the predicted reach of {\sc Ams-02}, taken from~\cite{AMSantiHe}. Although there might be other experiments which might have the capabilities of detecting anti-He, 
\footnote{Most notably the {\sc Gaps} experiment~\cite{GAPS} is aimed at the search of anti-deuterons with a dedicated technique which consists in slowing down the anti-deuteron entering the apparatus and then detecting the pion and X ray signatures of its annihilation on the nuclei in the matter of the detector itself. It is possible that such an analysis can be adapted to other anti-nuclei for the {\sc Gaps} set-up currently under development, with dedicated studies (Tsuguo Aramaki, private communication).}
we decide to limit the analysis to \AMS\ as a benchmark case. 

\medskip

The fluxes in the left panels in Fig.~\ref{fig:results} are obtained adopting our fiducial value for the coalescence momentum $p_{\rm coal} = 195$ MeV  (as well as the fiducial value $p_{\rm coal} = 167$ MeV for the background, see the Appendix). However, as emphasized in Sec.~\ref{sec:coalescence}, the actual value of $p_{\rm coal}$ is highly uncertain. Moreover, even the effective description of coalescence as based on this single energy-independent parameter can be questioned. 
We therefore recompute the spectra spanning different values of $p_{\rm coal}$, both for the DM signal and for the background. The results are shown in the right panels of Fig.~\ref{fig:results}, where we show how the fluxes allowed by anti-proton constraints are modified: the red area in the right plots reproduces the red area in the left plots, while the hashed region refers to $p_{\rm coal} = 300$ MeV (and $p_{\rm coal} = 250$ MeV for the background, adopting the same scale ratio). For case iii) we increase further the value of $p_{\rm coal}$ to 600 MeV  ($p_{\rm coal} = 500$ MeV for the background), in order to explore what it would take to skim the \AMS\ sensitivity region.

\medskip

\begin{figure}[t]
\begin{center}
\vspace{-1cm}
\includegraphics[width=0.47\textwidth]{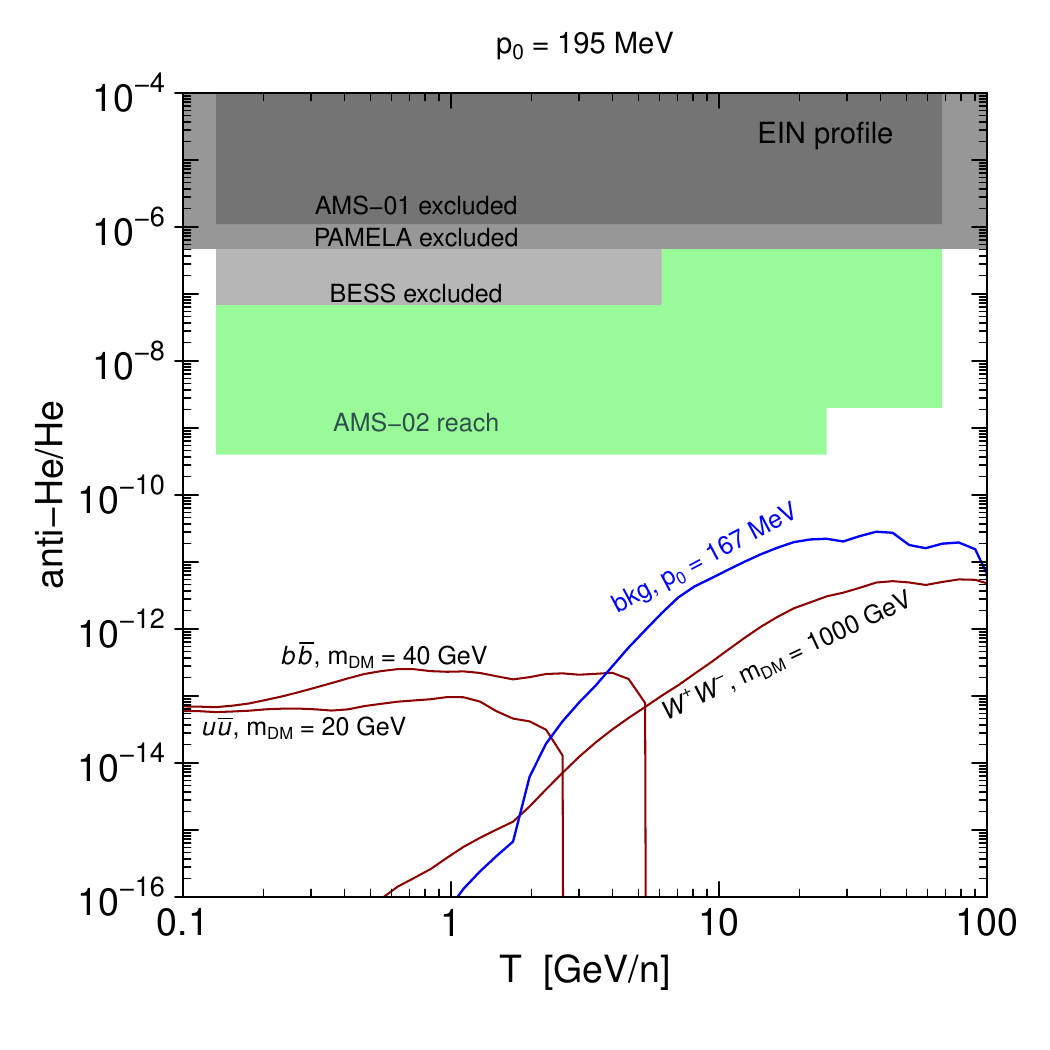}
\includegraphics[width=0.47\textwidth]{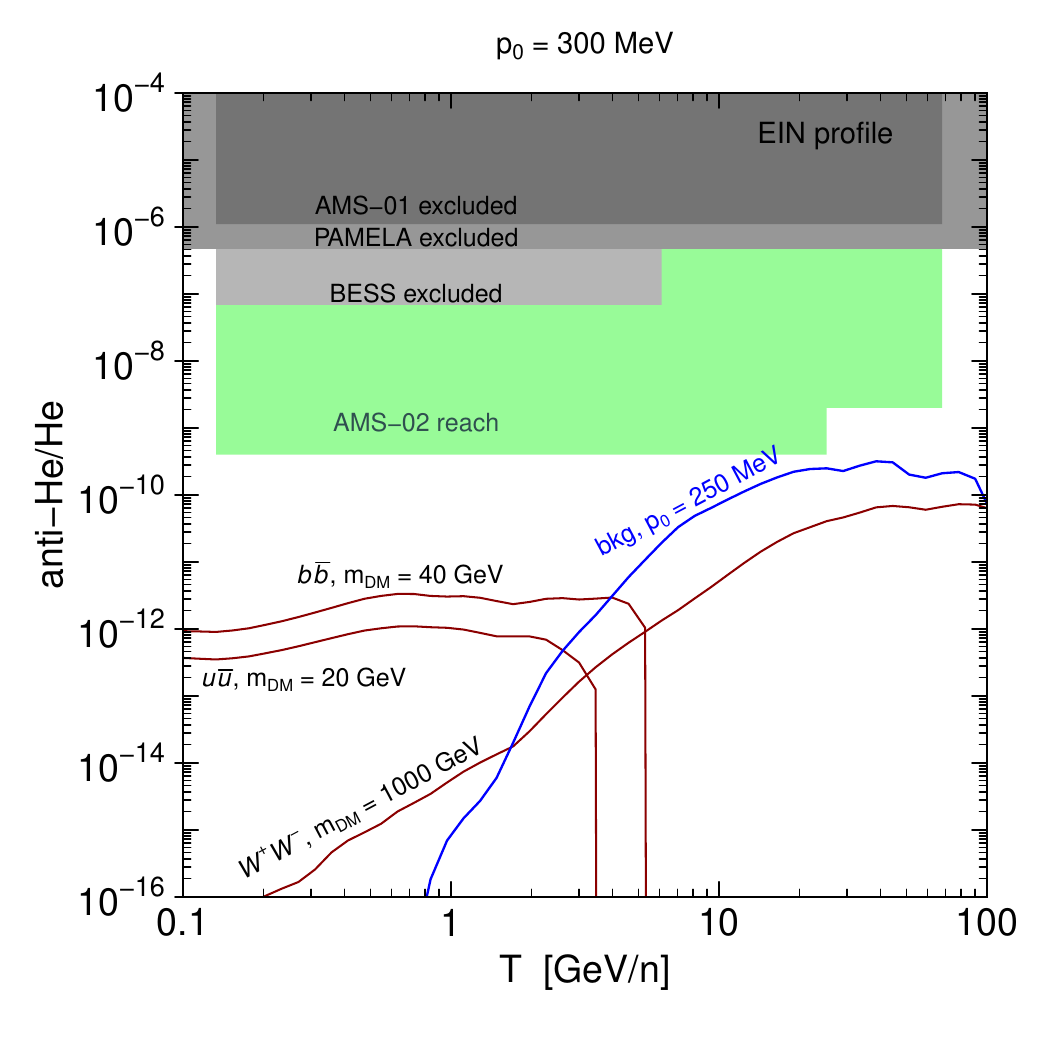} 
\caption{\small\em\label{fig:resultsratios} {\bfseries Predicted anti-He/He ratios} for the indicated DM benchmark models and for the largest flux admitted by anti-proton constraints. We recall that the annihilation cross section is taken as thermal for the light quarks and $b \bar b$ channels, while it is 10 times the thermal one for the $W^+W^-$ channel. The left plot assumes $p_{\rm coal} = 195$ MeV, the right one $p_{\rm coal} = 300$ MeV.}
\end{center}
\end{figure}

The inspection of the results in Fig.~\ref{fig:results} shows that, for all the cases that we have considered, the anti-He spectrum sits quite below the predicted reach of \AMS, so that the detection perspectives are rather dim. Fig.~\ref{fig:resultsratios} expresses the same information in terms of the anti-He/He ratio: with $p_{\rm coal}~=~195$ MeV, the DM signal reaches at most $\approx 10^{-13}$ in cases i) and ii) and $\approx 10^{-11}$ in case iii). These values increase by approximately one order of magnitude if $p_{\rm coal}~=~300$ MeV. They can be confronted with the \AMS ~expected sensitivity which is at the level of $10^{-9}$
~\cite{AMSantiHe}.

Increasing the DM annihilation cross section to augment the yield is not a viable possibility, given the stringent anti-proton constraints \cite{pbarconstraints}. The one example, among the ones we considered, in which the spectrum skims the \AMS\ sensitivity region (in the highest kinetic energy portion) is for the $W^+W^-$ channel, with a DM mass of 1 TeV and annihilation cross section 10 times larger than the thermal one (case iii, allowed by antiproton bounds), when we assume $p_{\rm coal} = 600$ MeV and if propagation is close to {\sc Max}. For this rather extreme case, however, somewhat unfortunately the shape of the spectrum resembles the one of the astrophysical background, such that, even in case of a positive detection of
anti-He nuclei, ascribing the events to a DM origin would be very challenging at best. On the positive side though, if the coalescence momentum for the background is as high as 500 MeV, as we assume in this example, the astrophysical contribution would lie within the reach of \AMS.


\section{Conclusions}
\label{sec:conclusions}

We have computed the production of anti-He nuclei ($\,^3\overline{\rm He}\,$) for DM annihilations in the galactic halo (performing, with {\sc Pythia}, a MC coalescence that fully takes into account the phase space correlations between the constituent anti-nucleons), computed their transport in the Galaxy (in the {\sc Min}, {\sc Med}, {\sc Max} framework) and determined the spectra at the top of the atmosphere at Earth. We focussed on a few specific DM model cases. We incorporated the constraints coming from anti-protons, showing how they restrict the available parameter space severely.

We found that the prospects for detection are currently rather weak, with the fluxes remaining from more than one to several orders of magnitude below the predicted reach of the \AMS\ experiment. It would take a very optimistic configuration of the annihilation, propagation and coalescence parameters to reach the \AMS\ sensitivity region.

While the search for antimatter in general, and exotic anti-nuclei in particular, remains a very interesting avenue for finally exposing a `smoking gun' signature of particle DM in the galactic halo, we find that for anti-He a much larger sensitivity or maybe a dedicated innovative experiment would be needed.


\medskip

\noindent {\bf Note Added.} 
While this work was being completed Ref.~\cite{Carlson:2014ssa} was posted on the arXiv. The two analyses are similar and reach the same qualitative conclusions. In our approach, we adopt a smaller value for the coalescence momentum and we do not sum the yield of the $\bar{p}\bar{p}\bar{n}$ coalescence channel, therefore obtaining more conservative estimates for the fluxes. We fully incorporate in the computation the stringent anti-proton constraints and explicitly show their impact. Finally, we compare the predicted fluxes with present antiHe bounds and with the sensitivity of current-generation experiments.

\medskip

\small
\paragraph{Acknowledgements}
We thank Massimo Gervasi, Rene Ong, Philip von Doetinchem and Tsuguo Aramaki for useful communications concerning {\sc Ams-02} and  {\sc Gaps}.
We also thank Paolo Panci for interesting discussions. 
This work is supported by the European Research Council ({\sc Erc}) under the EU Seventh Framework Programme (FP7/2007-2013) / {\sc Erc} Starting Grant (agreement n. 278234 - `{\sc NewDark}' project).
The work of M.C. is also supported in part by the French national research agency {\sc Anr} under contract {\sc Anr} 2010 {\sc Blanc} 041301 and by the EU ITN network {\sc Unilhc}. 
N.F. and A.V. are supported by the research grant {\sl Theoretical Astroparticle Physics} number 2012CPPYP7 under the program {\sc Prin} 2012  funded by the Ministero dell'Istruzione, Universit\`a e della Ricerca ({\sc Miur}) and by the research grant {\sl TAsP (Theoretical Astroparticle Physics)}
funded by the Istituto Nazionale di Fisica Nucleare ({\sc Infn}). N.F. and A.V. are also supported
by the  {\sl Strategic Research Grant: Origin and Detection of Galactic and Extragalactic Cosmic Rays} funded by Torino University and Compagnia di San Paolo. N.F. acknowledges support of the spanish {\sc Micinn} Consolider Ingenio 2010 Programme {\sc Multidark} CSD2009--00064. M.C., M.T. and A.V. acknowledge the hospitality of the Institut d'Astrophysique de Paris ({\sc Iap}) and of the Theory Unit of {\sc Cern} where part of this work was done.


\bigskip

\appendix

\section{Determination of the astrophysical background}
\label{sec:background}

The astrophysical background to the Dark Matter anti-He flux is the result of the spallation of primary (or secondary) cosmic rays impinging on the Hydrogen and Helium nuclei of the interstellar medium, which are at rest. Thus, this background flux is mainly the sum of the contributions of six different reactions: $p\,p$, $p$\,He, He\,$p$, He\,He, $\bar{p}\,p$ and $\bar{p}$\,He.  As shown in \cite{Ibarra_bkg} for the case of  anti-deuterons, the weights of these contributions are not the same in the different energy ranges but the flux produced by $p\,p$ collisions largely dominate over the others (apart from the extremely low energy tail, in which the contribution from the $\bar{p}\,p$ and $\bar{p}$\,He processes can be sizable). 

Our goal in this Appendix is to compute the flux of anti-He produced by $p\,p$ collisions, in the framework of the \emph{event-by-event} coalescence model that we have described in Section \ref{sec:coalescence} and that we have used for the computation of the DM signal.

\begin{figure}[t]
\begin{center}
\vspace{-1cm}
\includegraphics[width=0.47\textwidth]{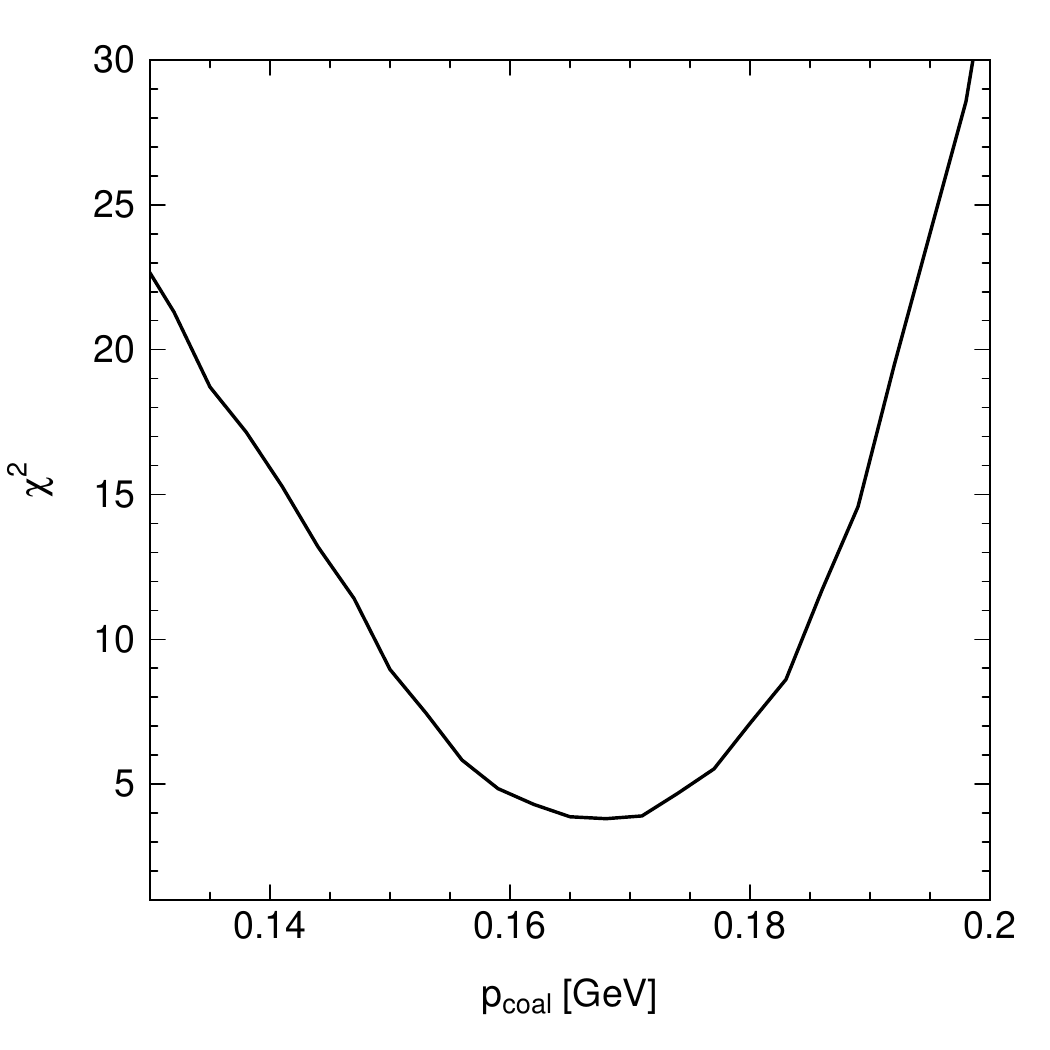}
\includegraphics[width=0.47\textwidth]{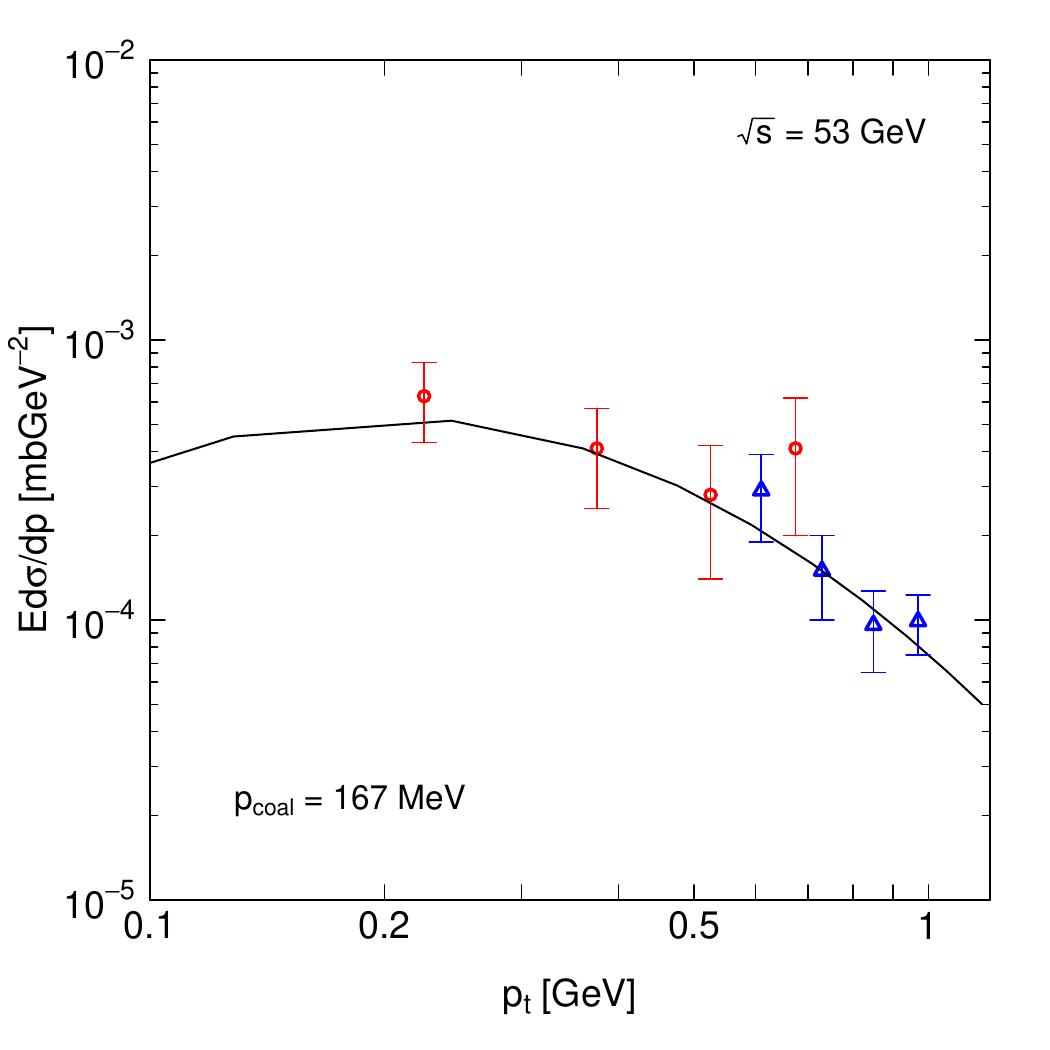} 
\caption{\small\em\label{fig:back_tuning} {\bfseries Tuning of the coalescence model (for the astrophysical background)}: in the left panel we report the total $\chi^2$ obtained by comparing the result of our model with the data in \cite{Alper1973265,Gibson} as a function of the coalescence momentum; in the right panel, the anti-deuteron production cross section for the best fit configuration is shown together with data points from \cite{Alper1973265} (in blue) and from \cite{Gibson} (in red).} 
\end{center}
\end{figure}

As already remarked for the DM case, we have to deal with the lack of experimental measurements on which to rely for the tuning of the coalescence momentum $p_{\rm coal}$:  to be consistent with the choices that we have made for the DM case, we choose to use for this parameter the value that is compatible with the measured cross section for the {\em anti-deuteron} production in similar (`fixed target') spallation processes. This observable  has been measured, for a center of mass energy $\sqrt{s}=53$ GeV, by the ISR experiment at CERN~\cite{Alper1973265,Gibson}. We report the results of our tuning process in Fig.~\ref{fig:back_tuning}: we find that, in order to reproduce the observed results, the correct value of $p_{\rm coal}$ to use is 167 MeV. This is then different from the value ($p_{\rm coal} = 195$ MeV) used as our benchmark choice for the DM signal. Naively, one would expect this parameter to be the same despite the initial state of the process, since the coalescence momentum is related to the probability for the $\bar{p}\bar{n}$ pair to merge into an anti-deuteron. Unfortunately, as stressed in \cite{Dal:2014nda}, the details of the hadronization process implemented through the Monte Carlo event generator can play a role and affect the phase space correlations between particles in the final state in different ways and thus a specific value of the coalescence momentum which can be good for a certain process can be not compatible with another. Therefore we deem more appropriate to apply the fitting procedure to the two processes separately and use two different values. Nevertheless, for illustration, we will later also show the computation of the background using the same value ($p_{\rm coal} = 195$ MeV) used for the DM signal.

Next, we need to solve the transport equation in the Galaxy. This is completely analogous to the formalism described in Section \ref{sec:propagation}, with the difference that, in the present case, the source term is
\begin{equation}
Q_{\rm sec} = \int_{E_{\rm thr}}^\infty dE' \Big( 4\pi \, \phi_p(E')\Big) \frac{d\sigma_{pp\rightarrow\overline{\rm He}+X}}{dE}(E,E') \ n_{\rm H},
\label{eq:bkg_source}
\end{equation}
where $E_{\rm thr} = 31 \ m_p$ is the threshold energy for the production of a single anti-He nucleus in a fixed target $p\,p$ collision, $\phi_p(E')$ is the primary proton flux and $n_{\rm H}$ is the Hydrogen nuclei density in the ISM (as discussed in Sec.~\ref{sec:propagation}, it consists of a thin slice of thickness $h=0.1\,{\rm kpc}$ and constant density 1 particle$/{\rm cm}^3$). The differential cross section is related to the energy distribution of the produced anti-He nuclei in the following way: 
\begin{equation}
\label{eq:source_sec}
\frac{d\sigma_{pp\rightarrow\overline{\rm He}+X}}{dE}(E,E') = \sigma_{pp,{\rm tot}}(E,E')\frac{dn_{\overline{\rm He}}}{dE}(E,E'),
\label{eq:bkg_source_2}
\end{equation}
being $n_{\overline{\rm He}}$ the number of anti-He nuclei in the energy range $[E,E+dE]$ normalized with respect to the total number of events. 
\begin{figure}
\begin{center}
\vspace{-1cm}
\includegraphics[width=0.5\textwidth]{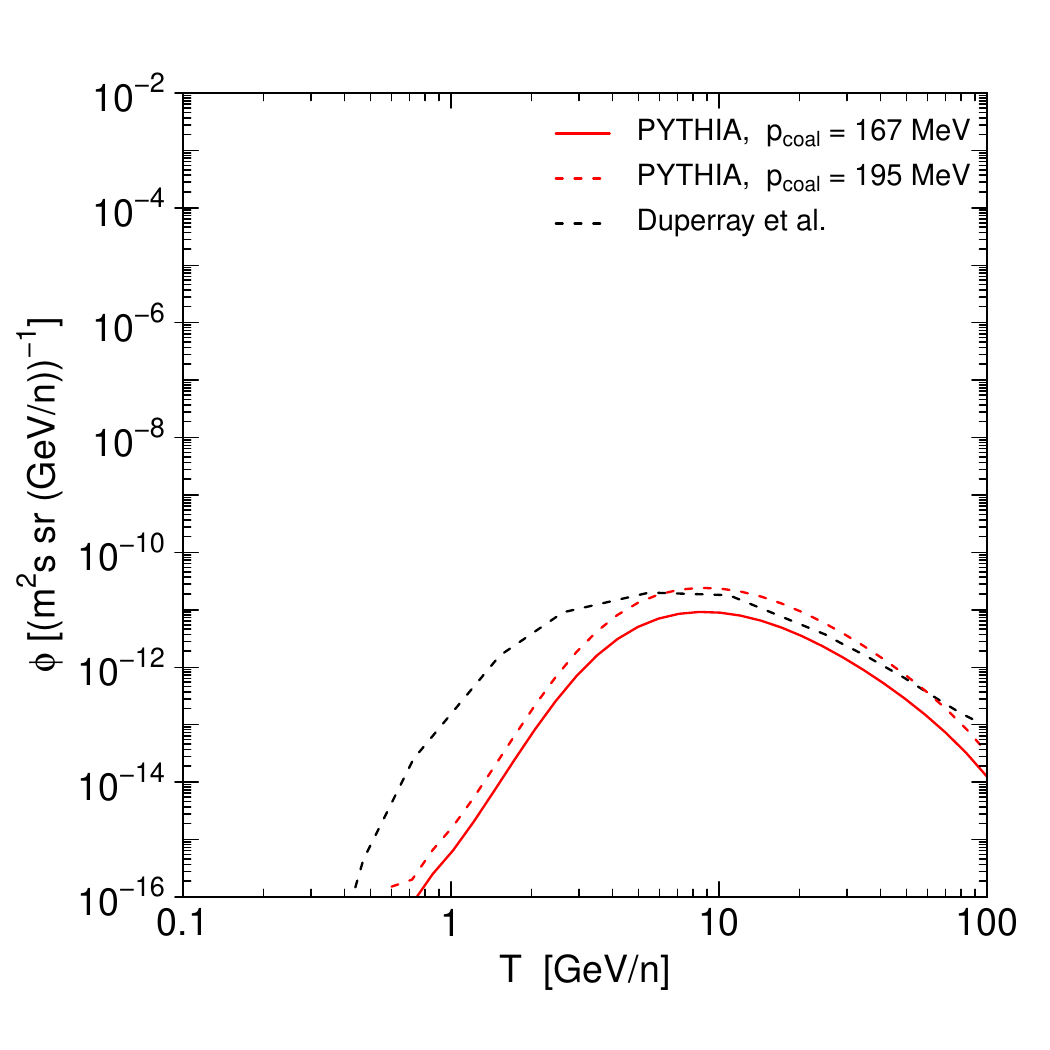}
\caption{\small\em\label{fig:background} {\bfseries Anti-He background fluxes}: our estimate for the anti-He background produced by $p\,p$ collisions is shown for two values of the $p_{\rm coal}$ parameter: 167 MeV (red solid line) and 195 MeV (red dashed line). The background previously computed in \cite{Duperray} (black solid line) is shown here for comparison.} 
\end{center}
\end{figure}
 We first evaluate the term in Eq.~(\ref{eq:bkg_source_2}) by simulating $p\,p$ fixed target collisions with {\sc Pythia} for a grid of values for the incoming proton energy, in the range from $E'=E_{\rm thr}$ to $E'=50$ TeV (which we have assumed as upper bound).  We then compute the integral in Eq.~(\ref{eq:bkg_source}) by using for the primary proton flux the parameterization
\begin{equation}
\phi_p~[(({\rm GeV}/n)~m^2~s~sr)^{-1}]= A\,E_{\rm kin}^{-\gamma}~[{\rm GeV}/n]
\end{equation}
with $A=11830.42$ and $\gamma=2.712$; this parameterization has been obtained as a global fit of the fluxes (above 10 GeV) measured by the most recent experiments ({\sc Ams02}, {\sc Pamela}, {\sc Jacee}, {\sc Atic2}, {\sc Cream}, {\sc Runjob})\cite{CRdata}. 
The rest of the propagation computation follows what already discussed in Sec.~\ref{sec:propagation}. In particular, we compute propagation functions, exactly analogous to the ones in Fig.~\ref{fig:RfunctionsHe3}, for the sets of parameters ({\sc Min}, {\sc Med}, {\sc Max}) in Table~\ref{tab:proparam} and we apply solar modulation as discussed in the main text.

\medskip

In figure \ref{fig:background} we show our final result for the background flux together with the one that had previously been derived in \cite{Duperray}. We see that the shape and normalization differ somewhat from the previous computation, but the important features of peaking at a few GeV/n and turning off below $\sim$1 GeV is confirmed, as expected on the basis of the kinematical considerations. The impact of choosing a $p_{\rm coal}$ parameter equal to the one applied for the DM signal is also shown. In this figure we have chosen the {\sc Med} parameters, but in Fig.~\ref{fig:results} we plot the three choices (the differences are barely visible).

\footnotesize
  

\end{document}